\documentclass{article}
\usepackage{graphicx}
\usepackage[latin1]{inputenc}
\usepackage[T1]{fontenc}
\usepackage[english]{babel}
\usepackage{amsmath}
\usepackage{amsfonts}
\usepackage{amssymb}
\usepackage{fancybox, framed, color}

\newcommand{\vt}{\vspace{0.15cm}}
\newcommand{\hs}{\hspace*{0.52cm}}
\newcommand{\bc}{\begin{center}}
\newcommand{\ec}{\end{center}}
\newcommand{\vd}{\vspace{0.1cm}}
\newcommand{\vp}{\vspace{0.4cm}}
\newcommand{\bmp}{\begin{minipage}}
\newcommand{\emp}{\end{minipage}}

\newtheorem{thm}{Theorem}
\newtheorem{lem}{Lemma}
\newtheorem{conj}{Conjecture}
\newtheorem{cor}{Corollary}
\newtheorem{defin}{\hs Definition}
\newtheorem{fait}{Claim}
\newtheorem{fact}{Fact}
\newtheorem{rem}{Remark}
\newtheorem{probl}{Problem}

\newcommand{\br}{\vd\begin{rem}\rm}
\newcommand{\er}{\end{rem}\vt}
\newcommand{\bt}{\vd\begin{thm}}
\newcommand{\et}{\end{thm}\vt}
\newcommand{\bln}{\vd\begin{lem}}
\newcommand{\eln}{\end{lem}\vt}
\newcommand{\bco}{\vd\begin{conj}}
\newcommand{\eco}{\end{conj}\vt}
\newcommand{\bcor}{\vd\begin{cor}}
\newcommand{\ecor}{\end{cor}\vt}
\newcommand{\bdn}{\vd\begin{defin}}
\newcommand{\edn}{\end{defin}\vt}
\newcommand{\bfn}{\vd\begin{fait}}
\newcommand{\efn}{\end{fait}\vt}
\newcommand{\bfac}{\vd\begin{fact}}
\newcommand{\efac}{\end{fact}\vt}
\newcommand{\bprobl}{\vd\begin{probl}}
\newcommand{\eprobl}{\end{probl}\vt}

\newcommand{\U}{U}
\newcommand{\V}{V}
\newcommand{\Q}{Q}
\newcommand{\pu}{u}
\newcommand{\pv}{v}
\newcommand{\aij}{a_j^i}
\newcommand{\bij}{b_j^i}
\newcommand{\cij}{c_j^i}
\newcommand{\pa}{a}
\newcommand{\pb}{b}
\newcommand{\pc}{c}

\newcommand{\pw}{w}
\newcommand{\aijp}{(a_j^i)'}
\newcommand{\bijp}{(b_j^i)'}
\newcommand{\cijp}{(c_j^i)'}
\newcommand{\Aij}{A_j^i}
\newcommand{\Bij}{B_j^i}
\newcommand{\Cij}{C_j^i}
\newcommand{\Aijp}{(A_j^i)'}
\newcommand{\Bijp}{(B_j^i)'}
\newcommand{\Cijp}{(C_j^i)'}
\newcommand{\C}{\mbox{\it H} }
\newcommand{\A}{A}
\newcommand{\Gij}{G_j^i}
\newcommand{\CCj}{CC_j}
\newcommand{\CC}{CC}
\newcommand{\X}{X}

\newcommand{\Y}{Y}
\newcommand{\Z}{Z}
\newcommand{\Xij}{X_j^i}
\newcommand{\Tij}{T_j^i}
\newcommand{\Yij}{Y_j^i}
\newcommand{\Zij}{Z_j^i}
\newcommand{\Di}{D_i}
\newcommand{\G}{G}
\newcommand{\lex}{\leq_{\it lex}}
\newcommand{\E}{E}

\newcommand{\D}{D}
\newcommand{\Zet}{\mathbb{Z}}
\newcommand{\cC}{\mathcal{C}}
\newcommand{\py}{w}
\newcommand{\yY}{W}
\newcommand{\R}{R}

\renewcommand{\Box}{\rule{1.5mm}{3mm}}
\begin{document}

\bc
{\Large\bf NP-hardness of sortedness constraints}
\vd
\vp

{Irena Rusu}
\vp

{\small\it L.I.N.A., Université de Nantes, 2 rue de
la Houssinière, BP 92208, 44322 Nantes, France}
\ec

\vp

\noindent\rule{\textwidth}{0.5mm}

\noindent {\bf Abstract}

In Constraint Programming, global constraints allow to model and solve many combinatorial problems.
Among these constraints, several sortedness constraints have been defined, for which propagation algorithms
are available, but for which the tractability is not settled. We show that the {\sf sort$(\U,\V)$}
constraint (Older et. al, 1995) is intractable (assuming P$\neq$NP) for integer 
variables whose domains are not limited to intervals. As a consequence,
the similar result holds for the {\sf sort$(\U,\V, P)$} constraint (Zhou, 1996). Moreover, the 
intractability holds even under the stability condition present in the recently 
introduced {\sf keysorting$(\U,\V,Keys,P)$} constraint (Carlsson et al., 2014), and requiring that the order of the variables with 
the same value in the list $\U$ be preserved  in the list $\V$. Therefore,
{\sf keysorting$(\U,\V,Keys,P)$} is intractable as well.

\noindent \rule{\textwidth}{0.5mm}


\section{Introduction}
Constraint programming systems support an increasing number of global constraints, {\em i.e.} constraints
for which the number of variables is arbitrary. Such constraints define an important search space,
that may be pruned using constraint propagation algorithms. Implementing a certain notion of consistency,
a propagation algorithm removes infeasible values from the domains of the variables, and its  
efficiency  is evaluated both with regard to its ability to limit the search space, and with regard to its running time.
Dealing with global constraints in general, that is, without fixing a constraint (or a set of constraints), is 
intractable \cite{bessiere2004complexity}. However, each constraint has its own complexity, which may
range from tractability at all levels of consistency, as for the {\sf alldifferent} constraint 
\cite{van2001alldifferent},  to intractability at relatively low levels of consistency, as for 
linear equations \cite{schulte2005bounds}.  

The tractability of a particular constraint is not always settled when the constraint is defined,
and this is the case for the sortedness constraints {\sf sort$(\U,\V)$} \cite{older1995getting}, {\sf sort$(\U,\V,P)$} 
\cite{zhou1996constraint} and {\sf keysorting$(\U,\V,Keys,P)$} \cite{sicstus}. Although  one or several of these  constraints 
are implemented  in  well-known systems like SICStus Prolog \cite{sicstus}, Gecode \cite{gecode} and  Choco \cite{choco},
as well as in the constraint modelling language MiniZinc \cite{nethercote2007minizinc}, 
their hardness is unknown.

In this paper we show that the intuitively simplest of these sortedness constraints, namely {\sf sort$(\U,\V)$},
is intractable (unless $P=NP$) even in the case where the domains of the variables in $\U$ are disjoint, 
and this leads to the intractability of {\sf sort$(\U,\V,P)$} and of  {\sf keysorting$(\U,\V,Keys,P)$}.

The organisation of the paper is as follows. In Section \ref{sect:CC}, we give the terminology and notations used
in the paper. In Section \ref{sect:basic}, we transform, using ideas from \cite{mehlhorn2000faster}, the search for a support of {\sf sort$(\U,\V)$} into a 
graph matching problem that we call {\sc SortSupport}, and we
show how to associate with each instance of the NP-complete problem {\sc Not-All-Equal 3SAT} an instance of {\sc SortSupport}.
The next section is devoted to the proof that our construction is a polynomial transformation \cite{garey1979computer}, implying the
NP-completeness of  {\sc SortSupport}. In Section \ref{sect:hardness}, we deduce hardness results about 
the three sortedness problems. Section \ref{sect:conclusion} is the conclusion.

\section{Constraints and consistency}\label{sect:CC}

In this paper, we deal with constraints over integer domains.
Given a variable $w$, we denote $Dom(w)$ its domain, which is assumed to be a
finite set of integers. When $Dom(w)$ is written as an interval $[l..r]$ (with integer $l,r$ such that $l\leq r$) 
or a union of intervals,  we understand that it contains only the integers in the (union of) 
interval(s), {\em i.e.}  $[l..r]$ is defined as $[l..r]:=\{d\in \Zet\, |\, l\leq d\leq r\}$.

A {\em constraint} $\cC$ is a couple $(\yY,\R)$, where $\yY=\{\py_1, \py_2, \ldots, \py_t\}$ is a set of variables
with associated domains $Dom(\py_i)$, and $\R$ is a $t$-ary relation over $\Zet$ (equivalently, 
a subset of $\Zet^t$). The constraint $\cC=(\yY,\R)$ is {\em satisfied} by a $t$-tuple $\delta=(\delta_1, \delta_2, \ldots, \delta_t)$  
assigning the value $\delta_i$ to variable $\py_i$, $1\leq i\leq t$, if  $\delta\in \R$. Denote 
$\mathcal{D}:=Dom(\py_1)\times Dom(\py_2)\times \ldots \times Dom(\py_t)$. If $\cC$ is satisfied by a 
$t$-tuple $\delta\in \mathcal{D}$, then $\delta$ is a {\em support} of $\cC$. 

A {\em constraint satisfaction problem} (or {\em CSP}) is defined as a set of variables with their associated 
 domains, and a set of constraints defined on subsets of the variable set. A {\em solution} of a CSP
 is an assignement of values from the associated domains to the variables that satisfies all the constraints.
In order to solve a CSP, constraints are successively used to prune the search space, with
the help of propagation algorithms that often seek to enforce various consistency properties, 
defined below (following \cite{choi2006finite}). A domain $\mathcal{D}$ is said {\em domain consistent} for the constraint 
$\cC=(\yY,\R)$  if, for each variable $\py_i$, $1\leq i\leq t$, and for each value $\delta_i\in Dom(\py_i)$, there is a 
support of $\cC$ assigning the value $\delta_i$ to $\py_i$. Domain consistency is a strong requirement, for which 
the following variants of bounds consistency are progressively weaker, but often very useful, alternatives. 

Denote $\inf_{\mathcal{D}}(\py_i)$ and $\sup_{\mathcal{D}}(\py_i)$ respectively the minimum and maximum value in $Dom(\py_i)$.
We say that a domain $\mathcal{D}$ is {\em bounds$(\mathbb{D})$ consistent} for $\cC$ if for each variable $\py_i$ and for each value
$\delta_i\in\{\inf_\mathcal{D}(\py_i),\sup_\mathcal{D}(\py_i)\}$ there exist integers $\delta_j$ with $\delta_j\in Dom(\py_j)$,
$1\leq j\leq t$ and $j\neq i$, such that $(\delta_1, \delta_2, \ldots, \delta_t)$ satisfies $\cC$.
We say that a domain $\mathcal{D}$ is {\em bounds$(\Zet)$ consistent} for $\cC$ if for each variable $\py_i$ and for each value
$\delta_i\in\{\inf_{\mathcal{D}}(\py_i),\sup_{\mathcal{D}}(\py_i)\}$ there exist integers $\delta_j$ with 
$\inf_{\mathcal{D}}(\py_j)\leq \delta_j\leq \inf_{\mathcal{D}}(\py_j)$,
$1\leq j\leq t$ and $j\neq i$, such that $(\delta_1, \delta_2, \ldots, \delta_t)$ satisfies $\cC$. Finally,
we say that a domain $\mathcal{D}$ is {\em bounds$(\mathbb{R})$ consistent} for $\cC$ if for each variable $\py_i$ and for each value
$\delta_i\in\{\inf_{\mathcal{D}}(\py_i),{\sup_\mathcal{D}}(\py_i)\}$ there exist real numbers $\delta_j$ with 
$\inf_\mathcal{D}(\py_j)\leq \delta_j\leq \inf_\mathcal{D}(\py_j)$,
$1\leq j\leq t$ and $j\neq i$, such that $(\delta_1, \delta_2, \ldots, \delta_t)$ satisfies $\cC$.

We now define the sortedness constraints:

\begin{itemize}
 \item the {\sf sort$(\U,\V)$} constraint, defined in \cite{older1995getting}, has variable set $\U\cup \V$,
where $\U=\{\pu_1, \pu_2, \ldots, \pu_n\}$ and $\V=\{\pv_1, \pv_2, \ldots, \pv_n\}$, 
and is satisfied by a $2n$-tuple of values assigned to the variables if and only if the variables in $\V$ are the sorted list of the 
variables in $\U$. The correspondence between the variables in $\U$ and
those in $\V$ is therefore a permutation. Propagation algorithms achieving bounds$(\Zet)$-consistency 
have been proposed in \cite{guernalec1997narrowing} and \cite{mehlhorn2000faster}.

\item  the {\sf sort$(\U,\V,P)$} constraint, defined in \cite{zhou1996constraint}, generalizes the 
{\sf sort$(\U,\V)$} constraint by adding a set $P$ of $n$ variables with domains included in $\{1, 2, \ldots, n\}$
in order to bring the permutation into the variable set of the constraint. This constraint
thus has variable set $\U\cup\V\cup P$ and is satisfied by a $3n$-tuple of values if and only if (a) the variables
in $\V$ are the sorted list of the variables in $\U$, (b) the variables
in $P$ are all distinct, and (c) the permutation associating the variables from $\U$ and $\V$ is
the one defined by the variables in $P$. The propagation algorithms for the {\sf sort$(\U,\V)$} constraint
are able to reduce the domains of the variables in $\U\cup\V$ similarly, but achieve bound($\Zet$)-consistency only
on the $\V$-domains, and not on the $\U$ and $P$-domains \cite{thiel2004efficient}.

\item the {\sf keysorting$(\U,\V,Keys,P)$} constraint (where $Keys$ is a positive integer), defined in \cite{sicstus}, allows 
to add two features with respect to {\sf sort$(\U,\V,P)$} : (a) each variable is a $h$-tuple 
($h\geq 1$ and integer, common to all variables), whose first $Keys$ elements 
form the sorting key of the variable, using lexicographic ordering;
and (b) the sorting has to be {\em stable}, {\em i.e.} any pair of variables with the same key value
must have the same order in $\U$ and in $\V$. The domain $Dom(z)$ of any variable $z$ from $\U\cup \V$ is thus a 
$h$-tuple of domains. When $Keys=1$, the lexicographic order of the keys is the 
classical order between integers, and thus {\sf keysorting$(\U,\V,1,P)$} is similar to 
{\sf sort$(\U,\V,P)$}, except that it requires the stability of the sorting.

\end{itemize}

Given two non-empty sets of integers $D$ and $E$, we write  $D\lex E$ whenever there exist values $d\in D$ and $e\in E$ such that
$d\leq e$. This is not (and is not intended to be) an order on sets, but allows to compare the domains 
of the variables with respect to the possibility to have a given order between the assigned values.

\section{Links between sortedness, graph matching and 3SAT}\label{sect:basic}

Consider two sets of variables $\U=\{\pu_i\,|\,  1\leq i\leq n\}$ and $\V= \{\pv_i\, |\, 1\leq i\leq n\}$,
with finite integer domains $Dom(\pu_i)$ and $ Dom(\pv_i)$, for all $i$, $1\leq i\leq n$. Following \cite{mehlhorn2000faster},
we define the {\em intersection graph} $\Gamma(\U,\V)$ of $\U$ and $\V$ as the bipartite graph with vertex set $\U\cup\V$ and
edge set $\{(\pu_i,\pv_j)\, | \, Dom(\pu_i)\cap Dom(\pv_j)\neq \emptyset\}$. A {\em matching} of $\Gamma(\U,\V)$
is an injective function $\sigma:\V'\subseteq \V\rightarrow \U$ such that $(\sigma(\pv),\pv)$ is an edge of $\Gamma$
for each $\pv\in \V'$.
We also use the notation $M=\{(\sigma(\pv),\pv), \pv \in \V'\}$ to designate the same matching. A matching 
$M$ {\em saturates} a vertex $x$ if there exists an edge in $M$ with endpoint $x$. We say that $M$  is a {\em perfect matching}
if it saturates all the vertices in $\Gamma(\U,\V)$. 

Denote $\Q_i=Dom(\sigma(\pv_i))\cap Dom(\pv_i)$, for each $\pv_i$ for which $\sigma(\pv_i)$ is defined.
Then testing whether  {\sf sort}$(\U,\V)$ has a support is equivalent to solving the following problem: 
\medskip

\noindent{\sc SortSupport}

\noindent {\bf Instance:} \hspace*{0.01cm} \bmp[t]{12.4cm} Two sets of variables $\U=\{\pu_i\, |\, 1\leq i\leq n\}$ 
and $\V= \{\pv_i\,|\,  1\leq i\leq n\}$, with finite integer domains $Dom(\pu_i)$ and $ Dom(\pv_i)$, for all $i$, $1\leq i\leq n$. \emp\vd

\noindent {\bf Question:} \hspace*{0.01cm} \bmp[t]{12.4cm} Is there a perfect matching $\sigma:\V\rightarrow \U$ of $\Gamma(\U,\V)$ such that
$\Q_i\neq \emptyset$ for all $i$ with $1\leq i\leq n$ and $\Q_1\lex \Q_2\lex \ldots \lex \Q_n$?\emp\vd
\medskip

We show that {\sc SortSupport} is NP-complete, and this even in the case where the domains
$Dom(\pu_i)$, $1\leq i\leq n$, are disjoint. To this end, given $\Gamma(\U,\V)$, a matching fulfilling
the required conditions is called a {\em sort-matching}. Note that the order $\pv_1,\pv_2, \ldots, \pv_n$
of the elements in $\V$ is important, since it defines the sort-matching.
\bigskip

We adapt the graph construction in \cite{rusu2008maximum}, and therefore use the same notations. 
The reduction is from the NP-complete problem {\sc Not-All-Equal 3SAT} \cite{garey1979computer}, the variant of 3SAT in
which each clause is required to have at least one true and at least one false literal.

Let $\C= \C_1\land \C_2 \land \ldots \land \C_k$ 
be an instance of {\sc Not-All-Equal 3SAT}, where each clause $\C_i$, $i=1, 2, \ldots, k$,
contains three literals from the set  $x_1, \overline{x_1}, x_2, \overline{x_2},$ \ldots, 
$x_p, \overline{x_p}$. We assume that, for each $j=1, 2, \ldots, p$, the literal  
$x_j$ occurs in the instance $\C$ as many times as $\overline{x_j}$ (otherwise, 
add to $\C$ an appropriate number  of clauses $(x_j\vee x_j\vee \overline{x_j})$ or 
$(x_j\vee \overline{x_j}\vee \overline{x_j})$). We note $occ(j)$ the total number of occurrences 
of $x_j$ in a clause, either as a {\em positive occurrence} ({\em i.e.} as $x_j$) or as a {\em negative
occurrence} ({\em i.e.} as $\overline{x_j}$).

We wish to build an instance $\U,\V$ of {\sc SortSupport} such that $\Gamma(\U,\V)$
consists of:

\begin{figure}[t]
\centering
\includegraphics[width=9cm]{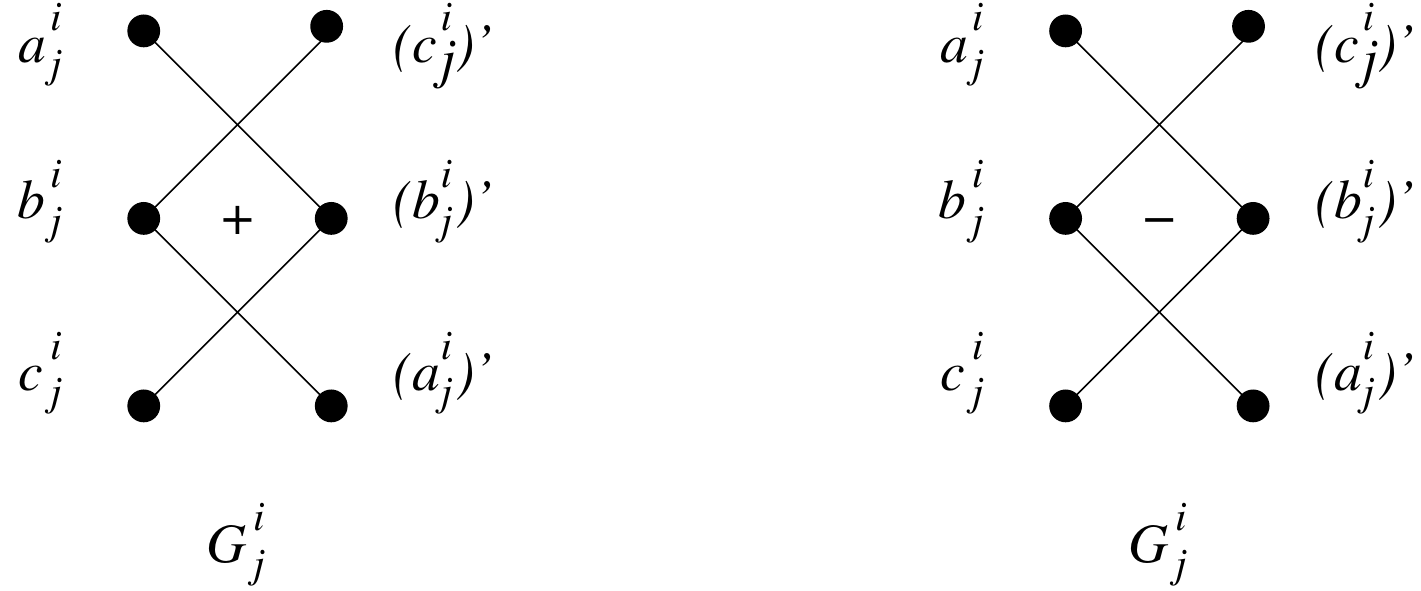}
\caption{\small  The unit graph $\Gij$ for a positive
or negative occurrence of  $x_j$ in  $\C_i$. Note that variables $\aij,\bij,\cij$ belong to $U$ and variables $\aijp,\bijp,\cijp$ 
belong to $V$.}
\label{fig:unit}
\end{figure}

\begin{figure}[t]
\centering
\includegraphics[width=14cm]{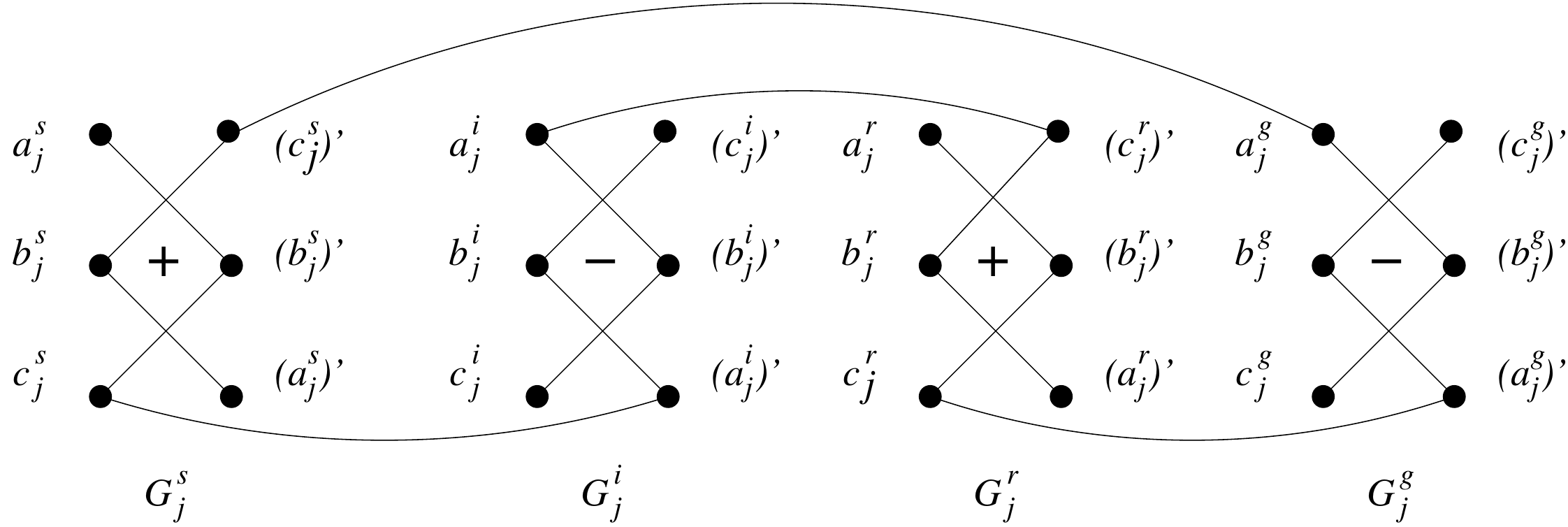}
\caption{\small The consistency component $\CC_j$ assuming that the literal
$x_j$ occurs in clauses $\C_s, \C_r$ and literal $\overline{x_j}$ occurs in clauses $\C_i, \C_g$. The horizontal edges are respectively the up-linking
edges and the  down-linking edges, according to their position.}
\label{fig:CCj}
\end{figure}

\begin{figure}[t]
\centering
\includegraphics[width=8.5cm]{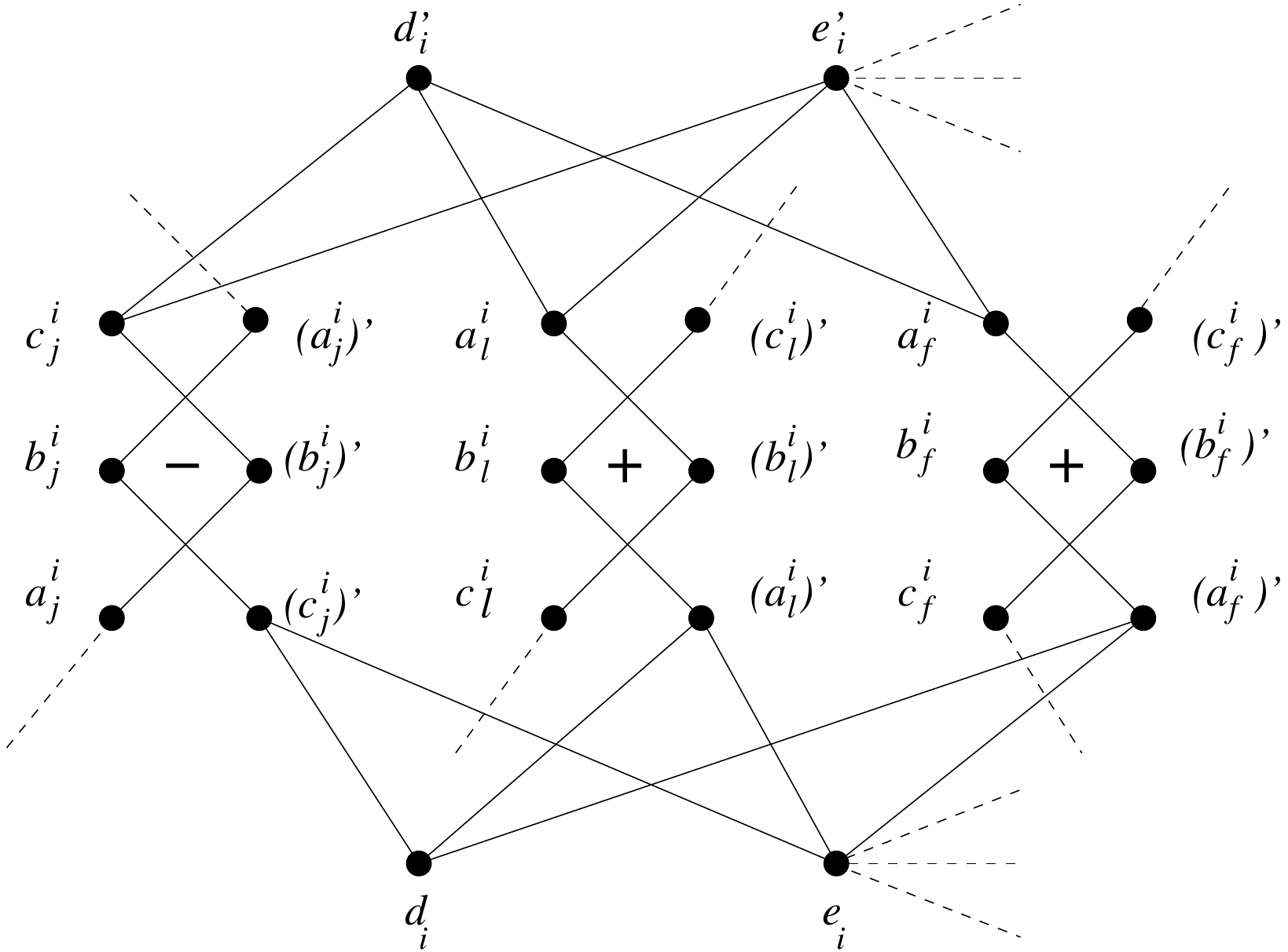}
\caption{\small The truth component $\Di$ assuming that the 
associated clause $\C_i$ is $(\overline{x_j}\vee x_l\vee x_f)$. The dashed lines outgoing from vertices belonging to unit graphs are 
the up- and down-linking edges inside the consistency component containing the vertex. The dashed lines outgoing from $e_i$ and $e'_i$
symbolize the set of edges in the completion component $\E$. Note that the
negative unit graphs are drawn with up-edges down, and vice versa.
}
\label{fig:Di}
\end{figure}

\begin{itemize}
\item a {\em unit graph} $\Gij$ (see Figure \ref{fig:unit}), for each positive or negative 
occurrence of a literal $x_j$ in a clause $\C_i$. The vertices of
$\Gij$ are the variables $\aij,\bij,\cij$ from $U$ and $\aijp,\bijp,\cijp$ from $\V$
joined by the four edges  $(\aij,\bijp)$,  $(\bij,\cijp)$ (called {\em up-edges}), 
$(\bij,\aijp)$ and $(\cij,\bijp)$ (called {\em down-edges}). Unit graphs are
positive or negative depending on the occurrence of $x_j$ they represent. 

\item  a {\em consistency component} $\CCj$ (see Figure \ref{fig:CCj}) for each literal $x_j$,
connecting all unit graphs associated with positive and negative occurrences of $x_j$. The unit
graphs are arbitrarily ordered such that they correspond alternately to a positive and
to a negative occurrence, the first unit graph being associated with a positive occurrence of $x_j$.
{\em Up-linking} edges join the $\aij$ vertex of a negative 
unit graph to the $(\pc_j^r)'$ vertex of the positive unit graph following it, in a circular way.
{\em Down-linking} edges join the $\cij$ vertex of a positive unit graph to the
$(\pa_j^r)'$ vertex of the negative unit graph following it.

\item  a {\em truth component} $\Di$ (see Figure \ref{fig:Di}) for each clause $\C_i$,
connecting the three unit graphs associated with the literals in $\C_i$. Note that the
negative unit graphs are drawn with up-edges down, and vice versa. Four vertices 
$d_i, e_i$ (defined to be in $\U$) and  $d'_i, e'_i$ (defined to be in $\V$) are added, 
the former ones joined to the 
$\cijp$ vertex of every negative unit graph and to the $\aijp$ vertex of every positive unit graph, whereas the latter
ones are joined to the $\cij$ vertex of every negative 
unit graph and to the $\aij$ vertex of every positive unit graph. These edges are
called {\em lateral edges}.

\item a {\em completion component} $\E$ providing, for each $i$ with $1\leq i\leq k$, 
an edge between $e_i$ and each $e'_j$ such that $i\neq j$.
\end{itemize}

The truth components allow to locally give truth values to the literals, whereas the
consistency components guarantee that the locally given truth values are globally
correct, that is, each literal is either true or false but not both. The completion component
ensures the existence of a perfect matching in the graph. Notice that, if a clause $\C_i$ contains 
two (or more) occurrences of the same literal $x_j$, then notations $\G_j^{i,1}$ and $\G_j^{i,2}$
should be used to identify the unit graph associated with each occurrence. We do not enter into
such details in our presentation, in order to keep it as simple as possible.

To define the required order on the elements in $\V$, we assume that in each unit graph the elements
$\aijp,\bijp,\cijp$ are ranged in this order, and in each consistency component $\CCj$ unit graphs are ordered
according to the alternate arbitrary order chosen to build $\CCj$ (recall that the first unit graph
in $\CCj$ is positive). Then,
the global order on $\V$ is built by considering the $\CC_1$, $\CC_2$, \ldots, $\CC_p$ 
consistency components in this order, and by adding vertices $d'_1, d'_2, \ldots, d'_k$,
$e'_1, e'_2, \ldots, e'_k$ in this order at the end.
\bigskip

{\bf Example.} Let $\C=\C_1\land \C_2$ with  $\C_1:\overline{x_1}\vee x_2\vee x_3$ and $\C_2:x_1\vee \overline{x_2}\vee \overline{x_3}$.
Then the consistency component $\CC_1$ contains the unit graphs $\G_1^2$ (positive) and $\G_1^1$ (negative), in this order; $\CC_2$ contains 
$\G_2^1$ (positive) and $\G_2^2$ (negative), in this order; and $\CC_3$ contains $\G_3^1$ (positive)
and $\G_3^2$ (negative), in this order. There are two truth components $D_1$ and $D_2$ corresponding
respectively to $\C_1$ and $\C_2$. By definition, $\D_1$ includes the unit graphs $\G_1^1, \G_2^1$ and $\G_3^1$, 
whereas $\D_2$ contains the unit graphs $\G_1^2, \G_2^2$ and $\G_3^2$.
The completion component has only two edges, $(e_2,e'_1)$ and $(e_1, e'_2)$. The order
of the variables in $V$ is $(\pa_1^2)'$, $(\pb_1^2)'$, $(\pc_1^2)'$, $(\pa_1^1)'$, $(\pb_1^1)'$, $(\pc_1^1)'$,
$(\pa_2^1)'$, $(\pb_2^1)'$, $(\pc_2^1)'$, $(\pa_2^2)'$, $(\pb_2^2)'$, $(\pc_2^2)'$,
$(\pa_3^1)'$, $(\pb_3^1)'$, $(\pc_3^1)'$, $(\pa_3^2)'$, $(\pb_3^2)'$, $(\pc_3^2)'$, $d'_1$, $d'_2$, $e'_1$, $e'_2$.

\bigskip

To finish our construction, we have to define the domains of the vertices in $\U$ and 
in $\V$ that exactly define the sought intersection graph $\Gamma(\U,\V)$. To simplify the 
notations, the domain of a vertex is denoted similarly to the vertex, but with an upper
case instead of a lower case, ({\em e.g.} $Dom(\aij)$ is denoted $\Aij$), except 
for the vertices $d_i, d'_i, e_i, e'_i$. 

\begin{figure}[t]
\centering
\scalebox{0.5}{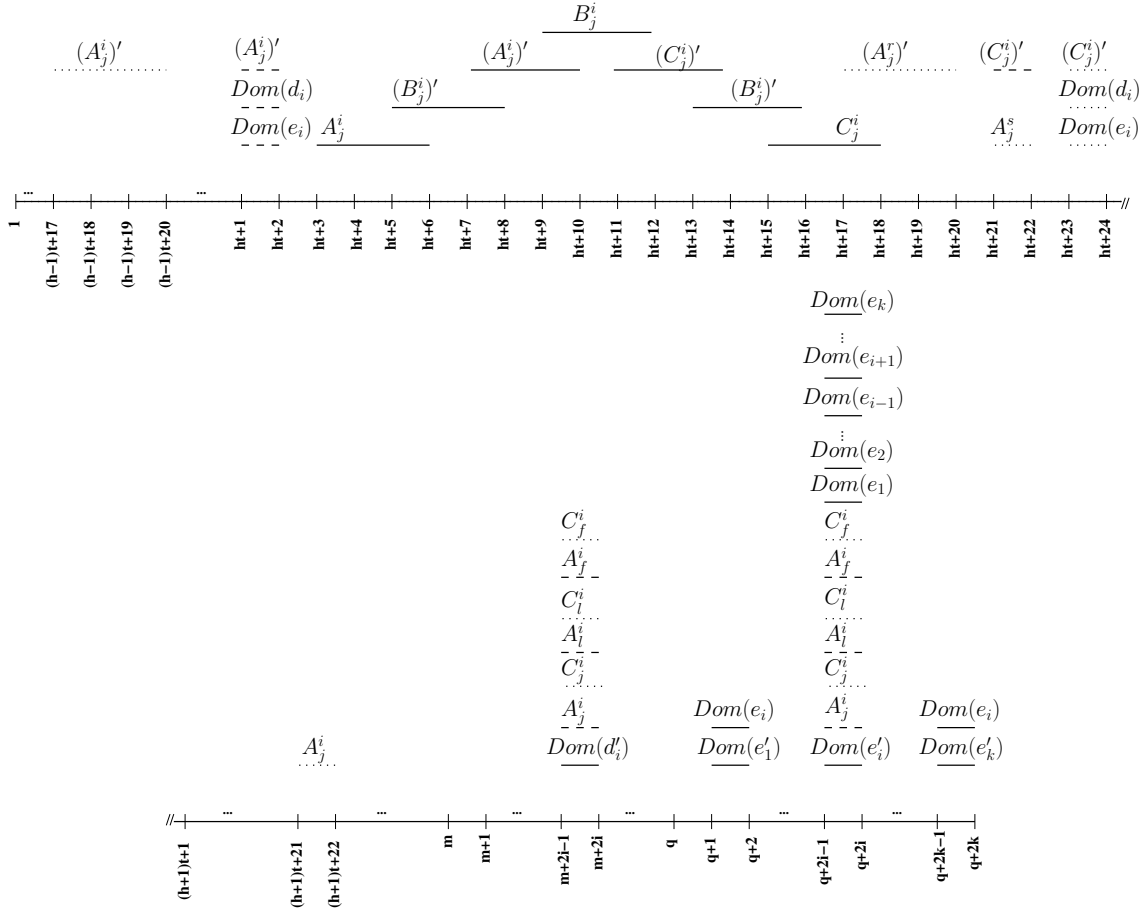}
\caption{\small Domains of the variables in $\Gij$ and $D_i$, for fixed $i$ and $j$, and their 
overlaps with other domains. The clause $\C_i$ is assumed to contain occurrences of literals
$x_j, x_l, x_f$. The notations $\A_j^s$  and $(\A_j^r)'$ refer respectively to the domains of the variables $\pa_j^s$
and $(\pa_j^r)'$, from the unit graphs $\G_j^s$ and $\G_j^r$ that precede and respectively  follow
$\Gij$ in the circular order of $\CCj$. The main domains of the variables are drawn with plain lines. 
Their secondary domains are drawn with dashed lines for a positive $\Gij$, and with dotted 
lines for a negative $\Gij$ (and this convention is extended to $\G_j^s$ and $\G_j^r$). We assumed 
that we are not in the third case of the definition of $\Xij$.
}
\label{fig:T}
\end{figure}

\br
Note that in the sequel we do not seek to minimize the sizes of the domains we define, as this is not important
for the proof of NP-completeness. In particular, we avoid domains that are singletons, in
order to allow a better illustration of the domains and their intersections in
Figure~\ref{fig:T}.
\er

%

Each unit graph (see Figure \ref{fig:T}) is defined by domains included into an interval 
of $\Zet$, of $t=24$ consecutive integers (that we call a  {\em block}),
so that each consistency component $\CCj$ is defined on $t*occ(j)$ consecutive integers,
and all the consistency components are represented on the interval $[1..t*\Sigma_{j=1}^p occ(j)]$. 
Noticing that $\Sigma_{j=1}^p occ(j)=3k$,
where $k$ is the number of clauses, and defining $m=3kt$, we define first the domains 
for the variables $d'_i$, $1\leq i\leq k$,  used in the truth
components, as follows: 
\medskip

$Dom(d'_i)=[m+2i-1.. m+2i]$
\medskip

\noindent This interval is devoted to creating the lateral edges in $\Di$.
Note that the last integer used by an interval in $Dom(d'_k)$ is 
$m+2k$, that we denote $q$. For each $i$, we define:
\medskip

$Dom(e'_i)=[q+2i-1.. q+2i]$.
\medskip

\noindent This interval is dedicated both to the lateral edges in $\Di$ and
to the edges in $\E$.

Now, each unit graph $\Gij$ is defined by domains inside the block $[ht+1..(h+1)t]$ (with little exceptions), where $h$ is the
number of unit graphs before $\Gij$ in the global order, {\em i.e.} $h=\Sigma_{q<j}occ(q)+r-1$,
where $r$ is the position of $\Gij$ in the alternate order of $\CCj$.
The domains of $\aij,\bij,\cij, \aijp$, $\bijp$, $\cijp$ are respectively defined as (see Figure \ref{fig:T}):
\medskip

$\Aij=[ht+3..ht+6]\cup \Xij$

$\Bij=[ht+9.. ht+12]$

$\Cij=[ht+15.. ht+18]\cup \Tij$

$\Aijp=[ht+7..ht+10]\cup \Yij$

$\Bijp=[ht+5..ht+8]\cup [ht+13..ht+16]$

$\Cijp=[ht+11..ht+14]\cup \Zij$
\medskip

\noindent where

\begin{equation*}
\Xij=\left\{
  \begin{array}{ll}
     Dom(d'_i)\cup Dom(e'_i),   & \mbox{if}\, \Gij\, \mbox{is a positive unit graph}  \\
     \left[(h+1)t+21.. (h+1)t+22)\right], & \mbox{if} \, \Gij\, \mbox{is a negative unit graph and, moreover,}\\
    & \Gij\, \mbox{is not the last unit graph in}\, \CCj \\
    \left[(h+1-occ(j))t+21.. \right. & \\
     \hspace*{1cm} \left. ..(h+1-occ(j))t+22\right], & \mbox{otherwise}
  \end{array}
\right.
\end{equation*}

\begin{equation*}
\Tij=\left\{
  \begin{array}{ll}
  \emptyset & \mbox{if} \, \Gij\, \mbox{is a positive unit graph}\\
     Dom(d'_i)\cup Dom(e'_i),  \hspace*{1.5cm} & \mbox{if}\, \Gij\, \mbox{is a negative unit graph} \hspace*{2.4cm}\\
    
  \end{array}
\right.
\end{equation*}

\begin{equation*}
\Yij=\left\{
  \begin{array}{ll}
    \left[ht+1..ht+2\right], \hspace*{2.2cm}  & \mbox{if}\, \Gij\, \mbox{is a positive unit graph} \hspace*{2.4cm} \\
    \left[ht-7.. ht-4 \right], & \mbox{if} \, \Gij\, \mbox{is a negative unit graph}\\
  \end{array}
\right.
\end{equation*}

\begin{equation*}
\Zij=\left\{
  \begin{array}{ll}
   \left[ht+21..ht+22 \right], & \mbox{if} \, \Gij\, \mbox{is a positive unit graph}\\
     \left[ht+23..ht+24\right],  \hspace*{2cm} & \mbox{if}\, \Gij\, \mbox{is a negative unit graph} \hspace*{2.2cm} \\
   
  \end{array}
\right.
\end{equation*}

The part of each domain defined by $\Xij, \Tij,\Yij$ or $\Zij$ before is called the {\em secondary domain} of
respectively $\aij, \cij, \aijp, \cijp$. The remaining part of the domain, {\em i.e.} the one explicitly stated
in the definitions of $\Aij, \Bij, \Cij, \Aijp, \Bijp, \Cijp$ above, is called the {\em main domain} of
the corresponding variables. Intuitively, the main domains allow to build the unit graphs and the down-linking edges of the 
consistency components. For a unit graph $\Gij$, the main domains of its variables belong
to its block $[ht+1.. (h+1)t]$. The secondary domains are devoted to the connections with
other unit graphs or with the other vertices of the truth components. 

It remains to give the domains of $d_i$ and $e_i$ for each $i$.  Recalling that a unit graph $\Gij$ exists 
if and only if $x_j$ has a positive or egative occurrence in $\C_i$, we define:
\medskip

$Dom(d_i)=\cup_{\Gij\, \mbox{is positive}} \Yij\cup \cup_{\Gij\, \mbox{is negative}} \Zij$

$Dom(e_i)=\cup_{\Gij\, \mbox{is positive}} \Yij\cup \cup_{\Gij\, \mbox{is negative}} \Zij\cup \cup_{s\neq i}Dom(e'_i)$
\medskip

\noindent Then, $Dom(d_i)$ contains intervals from the domains of the $\V$-variables in $\D_i$,
whereas $Dom(e_i)$ contains in addition the domains of all $e'_s$ with $s\neq i$.

\bigskip

The construction before, obviously polynomial, yields a bipartite graph with $n=11k$ vertices in each
part of the bipartition. Among these vertices $3k*6$ ($3k*3$ in each part) are in some
consistency component and $4k$ ($2k$ in each part) are in some truth component (but not 
in a consistency component). We show now that the intersection graph
of the variables we defined is indeed the graph we wished to build.

\bfn
The edges built using the domains 
defined above for the variables in $\U$ and $\V$ are exactly those of the unit graphs $\Gij$, consistency components $CC_j$,
truth components $\Di$ and completion component $\E$, for all $i$ and $j$ such that $x_j$ has a positive or negative
occurrence in $\Di$.
\efn

{\bf Proof.} We consider each variable and the edges it belongs to. It is easy to notice that the
main domain of a variable among $\aij, \bij, \aijp, \bijp$ or $\cijp$ can have non-empty 
intersections only with the main domains of the same set of variables, The only exception 
to this rule is $\cij$ (see below and Figure \ref{fig:T}). Moreover, the secondary domains of $\aij, \cij, \aijp, \cijp$
are specially defined to guarantee intersections with domains of other variables (according
to the desired edges in $\CCj$ and $\Di$), and thus have only empty intersections with 
the other domains in the block of $\Gij$. 

{\em Variable $\aij$.} The main domain of $\aij$ has non-empty intersection only with 
$\Bijp$ resulting into the edge $(\aij,\bijp)$. The secondary domain of  $\aij$
contains the domains of $d'_i$ and $e'_i$, if $\Gij$ is a positive unit graph, 
meaning that $\Aij\cap Dom(d'_i)\neq \emptyset$ and $\Aij\cap Dom(e'_i)\neq \emptyset$
and thus that the edges $(\aij, d'_i)$,  $(\aij, e'_i)$ exist. It is easy to see that no other edge
with endpoint $\aij$ exists in this case. If $\Gij$ is a negative unit graph,
then we have two sub-cases. If it is not the last unit graph in $\CC_j$, then $\Xij=[(h+1)t+21.. (h+1)t+22]$
and it has non-empty intersection only with the secondary domain  $\Z_j^r=[(h+1)t+21.. (h+1)t+22]$
of the variable $(\pc_j^r)'$ in the positive unit graph $\G_j^r$ immediately following it in $\CCj$. Then
we have the up-linking edge between $\Gij$ and  $\G_j^r$.
If $\Gij$  is the last unit graph in $\CC_j$, then $\Xij=[(h+1-occ(j))t+21.. (h+1-occ(j))t+22]$ and $\Xij$ is included
in the block of the first positive unit graph of $\CCj$, denoted $G_j^{i_1}$.
Then $\Xij$ has non-empty intersection only with the interval $\Z_j^{i_1}$ in the domain of the  variable $(\pc_j^{i_1})'$
of the first unit graph, and thus we have again the up-linking edge between the two unit graphs.

{\em Variable $\bij$.} Obviously, $\Bij$ has non-empty intersections with the main domains of $\Aijp$ and
$\Cijp$, defining the edges $(\bij,\aijp)$ and $(\bij, \cijp)$ of the unit graph. No other
non-empty intersections exist with $\Bij$.

{\em Variable $\cij$.} Again, we have an obvious intersection of $\Cij$ with $\Bijp$, implying 
the edge $(\cij,\bijp)$, and no other intersection with main domains. However, the main
domain $[ht+15.. ht+18]$ of $\cij$ also has non-empty intersection with the secondary domain
$\Y_j^r=[(h+1)t-7.. (h+1)t-4]$ of the variable $(\pa_j^{r})'$ in the unit graph following the current one,
if this latter graph is negative. We are therefore in the case where the two unit graphs
are joined by the down-linking edge $(\cij,(\pa_j^{r}))'$. Analyzing now the intersections of the
secondary domain of $\cij$, we have again two sub-cases. In the case $\Gij$ is negative, 
then its intersections with $Dom(d'_i)$ and $Dom(e'_i)$ are non-empty and we have the desired lateral edges in
the truth component for the clause $\C_i$. In the case $\Gij$ is
positive, there is no secondary domain for $\cij$.

{\em Variable $\aijp$.} On the main domain, $\Aijp$ has non-empty intersection with $\Bij$,
yielding the edge $(\bij, \aijp)$, and also with $\Bijp$, yielding no edge since both variables
are in $\V$. Consider now the secondary domain of $\aijp$. When $\Gij$ is positive, $\Yij$
is included into $Dom(d_i)$ and $Dom(e_i)$ and we have the sought lateral edges in $D_i$.
When $\Gij$ is negative, we have $\Yij=[ht-7..ht-4]$ which is
the same as $[(h-1)t+17.. (h-1)t+20]$, and thus has non-empty intersection with the main
domain of the vertex $\pc_j^s$ in the unit graph immediately preceding $\Gij$ in $\CCj$.
This confirms the down-linking edge already found for the variable $\pc_j^s$.

{\em Variable $\bijp$.} Non-empty intersections with domains of variables from $\U$
are found only for variables $\aij$ and $\cij$, confirming the edges already found above.

{\em Variable $\cijp$.} The edge $(\bij, \cijp)$ is confirmed by the non-empty intersection
of $\Cijp$ with $\Bij$. No other intersection is found with the main domain of $\cijp$.
For its second domain, we have  the intersection with $Dom(d_i)$ and $Dom(e_i)$ yielding the
corresponding lateral edges, in the case where $\Gij$ is negative. In the contrary case,
$\Zij=[ht+21..ht+22]$ and the only possible intersection is with some $\X_j^s$ defined
for another unit graph $\G_j^s$, which must be negative according to the definition of $\X_j^s$.  
With $\X_j^s=(h'+1)t+21, (h'+1)t+22]$, we must have $h'=h-1$, meaning that $\G_j^s$ is the
negative unit graph immediately preceding $\Gij$ in $\CCj$, and thus we have the
up-linking edge $(\pa_j^s,\cijp)$. With $\X_j^s=[(h'+1-occ(j))t+21.. (h'+1-occ(j))t+22]$,
we have that $\X_j^s$ is the secondary domain of the $\pa_j^{s}$  vertex in the last unit 
graph $\G_j^s$ in $\CCj$.  We also have that  $\X_j^s$ is included in the block 
of the first unit graph $\G_j^{i_1}$ in $\CCj$. Thus $\X_j^s$ has non-empty intersection with $\Zij$
if and only if $\Zij$ also concerns $\G_j^{i_1}$, that is if $\cijp=(\pc_j^{i_1})'$.
The resulting edge is then $(\pa_j^{s}, (\pc_j^{i_1})')$, which is the up-linking edge 
closing the circuit of unit graphs in $\CCj$.

{\em Variable $d_i$.} By definition, $Dom(d_i)$ is made of the three secondary domains 
of variables of type $\aijp$ and $\cijp$, according to the positive or negative
occurrence of $x_j$ in $\Di$, for which the edges have also been confirmed above.
We notice that all the other intersections with domains of variables from $\V$ are
empty.

{\em Variable $d'_i$.} By definition, $Dom(d'_i)$ has non-empty intersection
only with $\Xij$ (when $\Gij$ is positive) and with $\Tij$ (when $\Gij$ is negative)
and this yields the expected lateral edges.

{\em Variable $e_i$.} Again by definition, $Dom(e_i)$ is, as $Dom(d_i)$,
made of three secondary domains of variables of type $\aijp$ and $\cijp$
allowing to build the expected lateral edges, but also of $k-1$ domains of variables
$e'_i$ allowing to build the edges from $\E$ incident with $e_i$.

{\em Variable $e'_i$.} As all the edges and non-edges with vertices from $\U$
have already been verified, there is nothing more to check here.

All the edges in the unit graphs, consistency components, truth components and completion component
are correctly built,  and no undesirable edge is added. The claim is proved. $\Box$

\section{The proposed construction is a polynomial transformation}

In this section, we show that there is a truth assignment satisfying $\C$ with at least one
true and one false literal in each clause if and only
if $\Gamma(\U,\V)$ has a sort-matching. To this end, we first prove that:

\bfn
Let  $\Gij$ be an arbitrary unit graph. Then no sort-matching $M$ can contain simultaneously
the edges $(\aij,\bijp)$ and $(\bij,\aijp)$, nor simultaneously the edges $(\cij,\bijp)$ and $(\bij,\cijp)$.
\label{claim:simult}
\efn

{\bf Proof.} We have that $\Aij\cap \Bijp=[ht+5..ht+6]$, whereas $\Bij\cap\Aijp=[ht+9..ht+10]$.
Since $\aijp$ precedes $\bijp$ in the order on $\V$, if a sort-matching contained both edges $(\aij,\bijp)$ and $(\bij,\aijp)$
then we should have $\Bij\cap\Aijp\lex \Aij\cap \Bijp$,
and this is obviously false.

Similarly, $\Cij\cap \Bijp=[ht+15..ht+16]$, whereas $\Bij\cap\Cijp=[ht+11..ht+12]$,
which does not satisfy $\Cij\cap \Bijp \lex \Bij\cap\Cijp$. $\Box$

\bfn
Let $\CCj$ be an arbitrary consistency component. Then any sort-matching $M$ satisfies the following property:

a) either all the up-edges and all the down-linking edges in $\CC_j$ belong to $M$;

b) or all the down-edges and all the up-linking edges in $\CC_j$ belong to $M$.

\label{claim:2couplages}
\efn

\noindent{\bf Proof.} Given a unit graph $\Gij$ from $\CCj$, $M$ must contain
exactly one edge among $(\aij,\bijp)$ and $(\cij, \bijp)$, so that $\bijp$ is
saturated, and similarly exactly one edge among $(\bij, \aijp)$ and $(\bij, \cijp)$. 
By Claim \ref{claim:simult}, it results that $M$ contains either $(\aij,\bijp)$
and $(\bij, \cijp)$ (that is, the up-edges in $\Gij$) or $(\cij, \bijp)$ and $(\bij, \aijp)$
 (that is, the down-edges in $\Gij$). 
 
 In the former case, $\cij$ (if $\Gij$ is positive) or $\aijp$ (if $\Gij$ is negative) 
 can only be saturated by the down-linking edge with one endpoint in $\Gij$ (see Figure
 \ref{fig:CCj}), implying that the next (previous, respectively) unit graph $\G_j^{s}$ 
 in $\CCj$, in a circular way, also has its up-edges in $M$. The same deduction may be
 done for the unit graph $\G_j^{r}$ following (respectively preceding) $\G_j^{s}$, as
 follows. Since  the up-linking edge is not used by $M$ (the up-edges are already in $M$), 
 the vertex of  $\G_j^{r}$ incident
 with this edge must be saturated locally, and this can only be done by an up-edge of 
 $\G_j^{r}$. Thus, the property of a unit graph to have its up-edges and its incident
 down-linking edge in $M$ is propagated to all the consistency component $\CCj$.
 
 The reasoning is similar in the latter case. $\Box$
\bigskip

Now we are ready to prove the main result. For a truth component $\Di$ and a 
unit graph $\Gij$ in it, we call {\em $d'_i$-close} the up-edges of $\Gij$ if
$\Gij$ is positive, and the down-edges of $\Gij$ if $\Gij$ is negative. The
other edges in $\Gij$ are called $d_i$-close. In other words, given that in
the definition of $\Di$ the negative unit graphs are drawn upside down (see Figure \ref{fig:Di}),
the $d'_i$-close edges of $\Gij$ are the pairs of up- or down-edges one of whose endpoints is joined
to $d'_i$ (and similarly for $d_i$).

\bfn There is a truth assignment satisfying $\C$ with at least one true and one
false literal in each clause if and only if $\Gamma(\U,\V)$ admits a sort-matching.
\label{claim:main}
\efn

{\bf Proof.} To prove the {\em Only if} part, assume that $x_1, x_2, \ldots, x_p$
have been assigned boolean values satisfying $\C$ as required. Build  $M$, initially empty, as follows. 
For each clause $\C_i$, assume it contains the literals $x_j, x_l, x_f$,
with either a positive or a negative occurrence each. Assume without loss of generality,
that the (positive or negative) occurrence of $x_j$ is true and that the (positive or negative)
occurrence of $x_l$ is false. 

Then add to $M$:  the edges of $\Gij$ close to $d'_i$; the edges of $\G_l^i$ close to $d_i$;
the edges of $\G_f^i$ close to $d'_i$ or respectively to $d_i$ depending whether the occurrence
of $x_f$ is true or respectively false; the lateral edge joining $d'_i$ to $\G_l^i$
and the lateral edge joining $d_i$ to $\Gij$. Moreover, add to $M$ the up- or down-linking edges 
that are needed to saturate three of the four remaining unsaturated  vertices of $\Gij, \G_l^i$
and $G_f^i$. The remaining unsaturated vertex belonging to a unit graph of $D_i$ is
a vertex $y$ of $\G_f^i$. If the occurrence of $x_f$ in $\C_i$ is false, then this vertex
is either $\pa_f^i$ (when the occurrence is positive) or $\pc_f^i$ (when the
occurrence is negative), and is always adjacent to $e'_i$. If, on the contrary,
 the occurrence of $x_f$ in $\C_i$ is true, then this vertex is either $(\pa_f^i)'$ (when
 the occurrence is positive) or  $(\pc_f^i)'$ (when the occurrence is negative),
 and is always adjacent to $e_i$.  Then add $(y,e'_i)$
to $M$ if $(y,e'_i)$ is an edge, and add $(e_i,y)$ to $M$ in the contrary case,
leaving thus $e_i$ unsaturated when the occurrence of $x_f$ is false in $\C_i$,
and $e'_i$ unsaturated when the occurrence of $x_f$ is true in $\C_i$. Equivalently,
$e_i$ (respectively $e'_i$) remains unsaturated when $\C_i$ is oversupplied  of (respectively
true) literals.

Now, as each consistency component has the same number of positive and negative occurrences
of its corresponding literal, it results that there are $3k/2$ true literals 
and $3k/2$ false literals in $\C$. Therefore, the number of clauses that
are oversupplied of true literals is the same as the number of clauses that are oversupplied
of positive literals, namely $k/2$ clauses in each case. Consequently, in the completion
component $\E$ the unsaturated vertices induce a $k/2$-regular bipartite graph.
By Hall's theorem \cite{hall1935representatives}, this graph has a perfect matching $M'$, that we add to $M$.
The construction of the matching is now complete.

We first have to show that $M$ is correctly built. The construction implies that in every
unit graph either both up-edges or both down-edges are in $M$. Inside any consistency component,
all unit graphs are in the same case among these two cases. To see this, let $\Gij$ and 
$\G_j^l$ be two neighboring unit graphs and assume without loss of generality that
$\Gij$ is positive and $\G_j^l$ is negative, and also that $x_j$ is true, implying that
$\overline{x_j}$ is false (the other cases are similar). Then in $\Di$ the up-edges of $\Gij$ are put into $M$ since they
are $d'_i$-close, whereas in $\D_l$ the up-edges of $\G_j^l$ are put into $M$ since
they are $d_i$-close. Thus in $\CCj$ either all the up-edges or all the down-edges
are in $M$, together with the down-, respectively up-, linking edges by the definition of
$M$. Moreover, in each $\D_i$ and in $\E$ the matching is correctly built by definition. Thus
$M$ is correctly built.

Obviously, $M$ is a perfect matching. Recall that we associate with it a bijective function
$\sigma:\V\rightarrow \U$ such that $\sigma(\pv)=\pu$ if and only if $(\pu,\pv)\in M$. 
In order to show the inclusion property  between sets $\Q(\pv):=Dom(\sigma(\pv))\cap Dom(\pv)$ 
required by a sort-matching, we show that for each pair $\pv,\pw\in \V$ such that
$\pv$ precedes $\pw$ in the order on $\V$ we have  $\Q(\pv) \lex \Q(\pw)$. 

This deduction is based on the following seven affirmations:
\medskip

{\em A1.} for each unit graph $\Gij$, $\Q(\aijp)\lex \Q(\bijp)\lex \Q(\cijp)$. 
\medskip

Indeed, as shown above, we have two cases. In the case where the up-edges of $\Gij$
belong to $M$, we have that $\sigma(\bijp)=\aij$, $\sigma(\cijp)=\bij$. Moreover,
$\sigma(\aijp)\in\{d_i,e_i\}$ if $\Gij$ is positive, and  $\sigma(\aijp) =\pc_j^s$
where $\G_j^s$ precedes $\Gij$ in $\CCj$, if $\Gij$ is negative. Then, $\Q(\aijp)=\Yij=[ht+1..ht+2]$
 and respectively $\Q(\aijp)=[(h-1)t+17.. (h-1)t+18]$ (where $h$ defines the block of $\Gij$). See Figure \ref{fig:T}.
 As $\Q(\bijp)=[ht+5..ht+6]$ and $\Q(\cijp)=[ht+11..ht+12]$, the affirmation follows. In the
 case where the down-edges of $\Gij$ belong to $M$, we have that  $\sigma(\bijp)=\cij$, $\sigma(\aijp)=\bij$.
 Moreover, $\sigma(\cijp)=\pa_j^s$ where $\G_j^s$ (circularly) precedes $\Gij$ in $\CCj$, if $\Gij$ is positive,
 and $\sigma(\cijp)\in\{d_i,e_i\}$, if $\Gij$ is negative. We deduce that 
$\Q(\cijp)=\Zij=[ht+21..ht+22]$ and respectively $\Q(\cijp)=\Zij=[ht+23.. ht+24]$. As 
$\Q(\aijp)=[ht+9..ht+10]$ and $\Q(\bijp)=[ht+15..ht+16]$, the affirmation follows.
\medskip

{\em A2.} for each pair of consecutive unit graphs   $\Gij, \G_j^r$ (in this order) in the same consistency component $\CCj$, $Q(\cijp)\lex Q((\pa_j^r)')$. 
\medskip

As before, consider first the case where the up-edges of $\Gij$ belong
to $M$ and deduce in the same way that $\Q(\cijp)=[ht+11..ht+12]$ (where $h$ defines 
the block of $\Gij$). Then, the up-edges of $\G_j^r$ also belong to $M$, therefore 
$\Q((\pa_j^r)')=\Y_j^r=[(h+1)t+1..(h+1)t+2]$, if $\G_j^r$ is positive, and respectively 
$\Q((\pa_j^r)')=[ht+17.. ht+18]$, if $\G_j^r$ is negative. The affirmation is then proved.
Consider now the case where the down-edges of $\Gij$ belong
to $M$ and deduce as before that $\Q(\cijp)=[ht+21..ht+22]$ and respectively 
$\Q(\cijp)=[ht+23.. ht+24]$ when $\Gij$ is positive, respectively negative. We also have
in this case that $\Q((\pa_j^r)')=[(h+1)t+9..(h+1)t+10]$, proving the affirmation.

 \medskip

{\em A3.} for each pair of consecutive unit graphs $\Gij, \G_{j+1}^r$ (in this order) in two consecutive 
consistency components, $Q(\cijp)\lex Q((\pa_{j+1}^r)')$. 
\medskip

Notice that in this
case $\Gij$ is negative and $\G_{j+1}^r$ is positive, therefore as before we
have either $\Q(\cijp)=[ht+11..ht+12]$ or $\Q(\cijp)=[ht+23.. ht+24]$, depending whether the
up- or the down-edges of $\Gij$ belong to $M$. Similarly, we also
have $\Q((\pa_{j+1}^r)')=\Y_j^r=[(h+1)t+1..(h+1)t+2]$ or $\Q((\pa_{j+1}^r)')=[(h+1)t+9..(h+1)t+10]$.
In all four possible cases, the affirmation holds.
\medskip

{\em A4.} for the last unit graph $\G_p^i$ in the last consistency component $\CC_p$,
we have $Q((\pc_p^i)')\lex Q(d'_1)$. 
\medskip

With the same deductions as above and since
$\G_p^i$ is negative, we have that either $\Q(\pc_p^i)=[ht+11..ht+12]$ or $\Q(\pc_p^i)=[ht+23.. ht+24]$,
where $h$ defines the block of $\G_p^i$, {\em i.e.} $ht+24=m=3kt$ (with $t=24$) since we have
$3k$ literals in the $k$ clauses and thus $3k$ blocks, of which $\G_p^i$ uses the last one.
Now, $\Q(d'_1)=[m+1..m+2]$ since $Dom(d'_1)=[m+1..m+2]$ and it is included in the domains of the
three variables yielding vertices adjacent to $d'_1$ in $\D_1$, and the 
affirmation follows.
\medskip

{\em A5.} for each $i$ with $1\leq i<k$, we have $\Q(d'_i)\lex \Q(d'_{i+1})$. 
\medskip

This is obvious by
the definition of the domains and the observation that $\Q(d'_s)=Dom(d'_s)$
for all $s$ with $1\leq s\leq k$.

\medskip

{\em A6.} we have $\Q(d'_{k})\lex\Q(e'_1)$. 

\medskip 

Again, $\Q(d'_k)=Dom(d'_k)=[m+2k-1..m+2k]$ and similarly 
$Q(e'_1)=Dom(e'_1)=[q+1..q+2]$ where $q=m+2k$. The affirmation is proved.
\medskip

{\em A7.} for each $i$ with $1\leq i<k$, we have $\Q(e'_i)\lex \Q(e'_{i+1})$. 
\medskip

This is obvious by
the definition of the domains and the observation that $\Q(e'_s)=Dom(e'_s)$
for all $s$ with $1\leq s\leq k$.

Affirmations {\em A1-A7} allow to deduce that $M$ is a sort-matching.
\bigskip

We consider now the {\em If} part of the theorem. Assume therefore that a 
sort-matching exists in $\Gamma(\U,\V)$. By Claim \ref{claim:2couplages},
such a matching must contain, for each unit graph, either both its
low-edges or both its up-edges. Given that the matching is perfect,
we deduce that all vertices  $d_i$ and  $d'_i$ are saturated, and therefore that
in each $\Di$ there is a unit graph $\Gij$ whose $d'_i$-close edges are in
$M$ (the unit graph containing the vertex $y$ such that $(d_i,y)\in M$) and a unit graph 
$G_l^i$ whose $d_i$-close edges are in $M$ (the unit graph containing the 
vertex $z$ such that $(z, d'_i)\in M$). Call $\G_f^i$ the third unit graph
in $\Di$.

Define, locally to $\Di$, the (positive or negative) {\it occurrence} of
$x_j$ in $\Di$ to be true, that of $x_l$ to be false, and that of $x_f$
to be true if $M$ contains the edges of $\G_f^i$ close to $d'_i$,
respectively false if $M$ contains the edges of $\G_f^i$ close to $d_i$.
Deduce a local truth assignment for each literal $x_j, x_l$ and $x_f$.
Notice that:
\medskip

{\em A8.} The local truth assignment of the literal $x_r$, for each $r\in\{j, l, f\}$,
is true if and only if the up-edges of $\G_r^i$ belong to $M$.
\medskip

Indeed, we have that $x_r$ is true if and only if exactly one of the two
following cases occurs: either $x_r$ has a positive occurrence in $\Di$, in
which case this occurrence has been assigned the true value, so
the $d'_i$-close edges of $\G_r^i$ belong to $M$; or $x_r$ has a negative 
occurrence in $\Di$,  in which case $x_r$ has been assigned a true value 
if and only if its negative occurrence has been assigned a false value,
and this happens only if $M$ contains the  $d_i$-close edges of $\G_r^i$.
In the former case, the $d'_i$-close edges  are the up-edges of $\G_r^i$. 
In the latter case, due to the upside-down position of negative
unit graphs in $\Di$,  the  $d_i$-close edges are also the up-edges of $\G_r^i$.
And we are done.

Due to the affirmato  {\em A8} and to Claim \ref{claim:2couplages}, we deduce that all the 
local truth assignments are coherent. Moreover, each clause $\C_i$ has 
at least one true literal and one false literal, namely the occurrences
of $x_j$ and of $x_l$ (with the notations above). $\Box$

\section{Hardness of sortedness constraints}\label{sect:hardness}

The construction in the preceding section allows us to deduce the
NP-completeness of {\sc SortSupport} in general, but also in a
particular case of it, as follows.

\bfn
{\sc SortSupport} is NP-complete.
\efn

{\bf Proof.} Obviously {\sc SortSupport} belongs to NP. 
To show it is NP-complete, apply Claim~\ref{claim:main}. $\Box$
\bigskip

With a slight modification of the domains we defined for the variables, we also have that:

\bfn
{\sc SortSupport} is NP-complete, even in the case where the
domains $Dom(\pu_i)$, $1\leq i\leq n$, are pairwise disjoint.
\label{thm:disjoint}
\efn

{\bf Proof.} In order to have pairwise disjoint domains, the idea is: 
(a) to extend the intervals of $\Zet$ that are used by several 
sets $Dom(\pu)$ with $\pu\in \U$ (thus shifting to right all the other intervals 
so as to avoid unwished overlaps); (b)  to cut them into a sufficiently 
large number of  sub-intervals; and (c) to use a specific sub-interval for each
$Dom(\pu)$, thus insuring the disjointness without modifying the 
relative positions on the real line of the intervals defining the
sets $\Q(v)$.

The shared intervals are as follows: $\Yij=[ht+1..ht+2]$ (shared by $Dom(d_i)$ and $Dom(e_i)$,
when $\Gij$ is positive), $\Zij=[ht+23..ht+24]$ (shared by $Dom(d_i)$ and $Dom(e_i)$,
when $\Gij$ is negative), $Dom(d'_i)=[m+2i-1..m+2i]$ (shared by $\Xij$ when $\Gij$ is positive,
and by $\Tij$ when $\Gij$ is negative) and $Dom(e'_i)=[q+2i-1..q+2i]$ (shared by $\Xij$ when 
$\Gij$ is positive, by $\Tij$ when $\Gij$ is negative and by $Dom(e_s)$ with 
$s\neq i$). As an example, consider $\Yij$ in the case of a positive
unit graph $\Gij$. As $\Yij$ is shared by  $Dom(d_i)$ and $Dom(e_i)$,
it should be extended to an interval of length three ({\em e.g.}
 $[ht+1..ht+4]$ instead of $[ht+1..ht+2]$ as it is now), in which case 
 $Dom(d_i)$ and $Dom(e_i)$ would be affected a sub-interval each ({\em e.g.}
 $[ht+1..ht+2]$ and $[ht+3..ht+4]$ respectively). Of course, in this case, 
 the domain $\Aij$ of $\aij$ should start at $ht+5$ instead of $ht+3$
 (and similarly for the other domains), so as to avoid overlaps. This
 would result into an augmentation of the size $t$ of each block. 
 
 The proof of the correctness is very similar to the one above. The main difference is that
 some intersections between domains are slightly shifted. $\Box$
 
 \bt
 Testing whether {\sf sort$(\U,\V)$}, {\sf sort$(\U,\V,P)$} or {\sf keysorting$(\U,\V,Keys,P)$}
 has a support is NP-complete, even in the case where the variables in $\U$ have pairwise
 disjoint domains.
 \label{thm:supp}
 \et

 {\bf Proof.} For {\sf sort$(\U,\V)$}, the affirmation follows immediately by the equivalence
to {\sc SortSupport} noticed in Section \ref{sect:basic}, and by Claim \ref{thm:disjoint}.
Furthermore, {\sf sort$(\U,\V)$} is the variant of {\sf sort$(\U,\V,P)$} where 
each variable in $P$ has the domain $\{1, 2, \ldots, n\}$. As testing whether {\sf sort$(\U,\V,P)$}
has a support is obviously in NP, the previous remark allows to deduce the NP-completeness 
of the problem. Finally, {\sf sort$(\U,\V,P)$} and {\sf keysorting$(\U,\V,1,P)$} are
equivalent when the domains of the variables in $\U$ are pairwise disjoint, since the stability
of the sorting is trivially satisfied by any assignment of values to the variables. $\Box$


\bigskip

Given a constraint $\cC$ defined as in Section \ref{sect:CC}, enforcing domain consistency requires 
to test whether for a given variable $\py_i$ and a given value $\delta_i\in Dom(\py_i)$, a 
support of $\cC$ exists assigning the value $\delta_i$ to $\py_i$. We can easily deduce that:

\bcor
Enforcing domain consistency for each of the constraints  {\sf sort$(\U,\V)$}, {\sf sort$(\U,\V,P)$} and 
{\sf keysorting$(\U,\V,Keys,P)$} is intractable, even in the case where the variables in $\U$ have pairwise
 disjoint domains.
\ecor

{\bf Proof.} By contradiction and for each of the three constraints, assume a polynomial algorithm $\mathcal{A}$ exists 
for testing the existence of a support with a given value for a given variable.  Recall that, by Claim \ref{claim:main},
an instance $\C$ of {\sc Not-All-Equal 3SAT} is satisfiable if and only if $\Gamma(\U,\V)$ admits a sort-matching,
and this latter affirmation holds if and only if {\sf sort}$(\U,\V)$ has a support. By applying $\mathcal{A}$
to all the four values in $\Bij$ for an arbitrarily  chosen variable $\bij$, we test in polynomial time
whether {\sf sort$(\U,\V)$} has a support. Then we have a polynomial algorithm for solving {\sc Not-All-Equal 3SAT},
a contradiction. The results for {\sf sort$(\U,\V,P)$} and {\sf keysorting$(\U,\V,Keys,P)$} easily follow. $\Box$ 
\bigskip

Focusing now on enforcing bounds consistency, we need to test whether for a given variable $\py_i$ and a given value 
$\delta_i\in\{\inf_{\mathcal{D}}(\py_i),\sup_{\mathcal{D}}(\py_i)\}$,
a $t$-tuple $(\delta_1, \delta_2, \ldots, \delta_t)$ satifsfying $\cC$ exists whose values $\delta_j$, $j\neq i$
are more or less constrained. More precisely, $\delta_j$ must belong to $Dom(\py_j)$, respectively to 
$[\inf_{\mathcal{D}}(\py_j)..\sup_{\mathcal{D}}(\py_j)]$, and
respectively to $[\inf_{\mathcal{D}}(\py_i),\sup_{\mathcal{D}}(\py_i)]$ to allow bounds$(\mathbb{D})$, respectively bounds$(\mathbb{Z})$
and respectively bounds$(\mathbb{R})$ consistency. We are able to show that:

\bt
Enforcing bounds$(\mathbb{D})$ consistency for each of the constraints  {\sf sort$(\U,\V)$}, {\sf sort$(\U,\V,P)$} and 
{\sf keysorting$(\U,\V,Keys,P)$} is NP-complete.
\label{thm:bounds}
\et

{\bf Proof.} It is easy to notice that these problems are in NP.
We show that the reduction from {\sc Not-All-Equal 3SAT} to {\sc SortSupport} in Section \ref{sect:basic}
allows to deduce the result for {\sf sort$(\U,\V)$}. Then the other results follow.

In the instance of {\sc SortSupport} built in Section  \ref{sect:basic}, the variable $d'_i$ 
has the domain $Dom(d'_i)=[m+2i-1..m+2i]$, so that $\inf_D(d'_i)=m+2i-1$ and $\sup_D(d'_i)=m+2i$.
Let $y=m+2i-1$ and let us notice that $\Gamma(\U,\V)$ has a sort-matching if and
only if it has a sort-matching such that $y\in \Q(d'_i)$. This is an easy consequence
of the observation that $Dom(d'_i)$ has non-empty intersection with another domain if
and only if it is included in it, {\em i.e.} if and only if $y$ belongs to the intersection.
Then by Claim \ref{claim:main}, we deduce the NP-completeness of testing whether there
is a support of {\sf sort$(\U,\V)$} assigning to $d'_i$ the value $y$. $\Box$

%
 \bigskip

Notice that the previous result is not proved for pairwise disjoint domains of variables in $\U$.
The reason is that in this variant the domain of $d'_i$ strictly overlaps the domains of other
variables and the proof of Theorem \ref{thm:bounds} is no longer valid.

\section{Conclusion}\label{sect:conclusion}
In this paper we have shown that the three sortedness constraints defined up to now are intractable,
even in the particular case where the variables to sort have pairwise disjoint domains, and even if 
we do not seek domain consistency but only enforcing bounds$(\mathbb{D})$ consistency. The
tractability of the lower levels of bounds consistency, {\em i.e.} bounds$(\Zet)$ and
bounds$(\mathbb{R})$ consistency, is shown for {\sf sort$(\U,\V)$} \cite{guernalec1997narrowing,mehlhorn2000faster},
but is still open for {\sf sort$(\U,\V,P)$} and {\sf keysorting$(\U,\V,Keys,P)$}.

\bibliographystyle{plain}
\bibliography{Contrainte}

\end{document}